\documentclass[onecolumn]{emulateapj}
\usepackage{graphicx}

\begin{document}

\def\etal{et al.\ \rm}
\def\ba{\begin{eqnarray}}
\def\ea{\end{eqnarray}}
\def\etal{et al.\ \rm}
\def\sgn{\mbox{sgn}}
\def\rmk{\mbox{k}}
\def\Bi{\mbox{Bi}}
\def\Ai{\mbox{Ai}}

\title{The origin of the negative torque density in 
disk-satellite interaction.}

\author{Roman R. Rafikov\altaffilmark{1,2} \& Cristobal 
Petrovich\altaffilmark{1}}
\altaffiltext{1}{Department of Astrophysical Sciences, 
Princeton University, Ivy Lane, Princeton, NJ 08540; 
rrr@astro.princeton.edu}
\altaffiltext{2}{Sloan Fellow}


\begin{abstract}
Tidal interaction between a gaseous disk and a massive orbiting 
perturber is known to result in angular momentum exchange between them. 
Understanding astrophysical manifestations of this coupling such as 
gap opening by planets in protoplanetary disks or clearing of gas by 
binary supermassive black holes (SMBHs) embedded in accretion disks 
requires knowledge of the spatial distribution of the torque
exerted on the disk by a perturber. Recent hydrodynamical 
simulations by Dong et al (2011) have shown evidence for the tidal 
torque density produced in a uniform disk to change sign at the radial 
separation of $\approx 3.2$ scale heights from the perturber's 
orbit, in clear conflict with the previous studies. 
To clarify this issue we carry out a linear 
calculation of the disk-satellite interaction putting special 
emphasis on understanding the behavior of the perturbed fluid 
variables in physical space. Using analytical as well as numerical 
methods we confirm the reality of the negative torque density 
phenomenon and trace its origin to the overlap of Lindblad 
resonances in the vicinity of the perturber's orbit --- an 
effect not accounted for in previous studies. These results 
suggest that calculations of the gap and cavity opening in disks 
by planets and binary SMBHs should rely on more realistic
torque density prescriptions than the ones used at present.
\end{abstract}



\section{Introduction.}  
\label{sect:intro}


Understanding the response of a gaseous disk to a periodic 
non-axisymmetric gravitational perturbation is an important 
astrophysical problem. It arises in a variety of contexts
such as the disk-planet interaction in protoplanetary disks, 
orbital evolution of supermassive black hole (SMBH) binaries 
embedded in circumbinary disks, dynamics of accretion disks 
in cataclismic variables, neutron star and black hole 
binaries, and so on.

Gravitational coupling between a gaseous disk and a massive
orbiting 
perturber inevitably leads to the angular momentum exchange 
between the two. This has two important 
consequences for the evolution of the system. First, the orbit
of the perturber can change leading to planet migration in
protoplanetary disks (Ward 1986) and to the black hole inspiral 
in the case of SMBH binaries embedded in gaseous 
disks (Ivanov \etal 1999; Gould \& Rix 2000). Second, the angular
momentum lost by the perturber gets absorbed by the disk, which 
causes redistribution of the disk surface density. In the
most dramatic cases, when the perturber is massive enough,
gas may be completely depleted in some parts of the disk, resulting
in gap opening in protoplanetary disks (Ward 1997) and cavity 
formation around the SMBH binaries (Armitage \& Natarajan 2002).

It is important to note that the latter process directly affects 
the former since the strength of tidal coupling between the disk 
and the perturber leading to migration depends on the spatial 
distribution of the disk surface density (Ward 1997). Thus,
the change of the disk properties caused by the gravitational 
influence of a perturber has an immediate and important effect 
on the orbital evolution of the perturber itself.

Disk-planet\footnote{From now on we will 
refer as ``disk-planet'' or ``disk-satellite'' to any other 
type of system where the 
tidal coupling is present, e.g. a SMBH binary surrounded by the 
disk, an accreting white dwarf-main sequence star binary, etc.} 
interaction leads to the excitation of nonaxisymmetric density 
waves, which carry the angular momentum lost by the perturber away 
from their excitation site. Subsequent dissipation of the waves
due to some linear or nonlinear process transfers this angular 
momentum to the disk material, ultimately driving evolution of 
the disk. The torque density (torque per unit radial distance) 
exerted on the disk material by damping waves $dT/dr|_d$ 
(i.e. the amount of angular momentum transferred by dissipating 
waves to the disk fluid per unit time and per unit radial 
distance) can thus be represented as
\ba
\frac{dT}{dr}\Bigg|_d={\cal L}*\left(\frac{dT}{dr}\right),
\label{eq:dTdr_d}
\ea
where $dT/dr$ is the {\it excitation torque density} --- the amount
of angular momentum {\it added} to the density wave per unit time and 
per unit radial distance by planetary tide. Operator ${\cal L}$
describes the wave damping --- transfer of the wave angular 
momentum to the disk material by some dissipation mechanism. This
operator is intrinsically nonlocal since the amount of angular 
momentum transferred by the wave to the disk at some location 
depends on the amount of angular momentum accumulated by the wave 
prior to that point. 

Thus, to know how the tidal coupling affects the disk one must 
understand both the {\it excitation} of the density waves 
(i.e. the radial dependence of $dT/dr$) and the details of their 
{\it damping} (i.e. the explicit form of operator 
${\cal L}$ in equation (\ref{eq:dTdr_d})). This important point 
was previously highlighted by Lunine \& Stevenson (1983), 
Greenberg (1983), Goldreich \& Nicholson (1989). The behavior of 
${\cal L}$ has been previously explored for linear viscous 
dissipation by Takeuchi \etal (1996) and for nonlinear damping by
Goodman \& Rafikov (2001) and Rafikov (2002). The results of these 
studies suggest that for low-mass perturbers wave dissipation is 
a nonlocal process and most of the wave damping occurs far 
from its driving region.

In this work we concentrate on the wave excitation and do not 
consider its propagation and dissipation. Driving of the density 
wave is characterized by the dependence of torque density $dT/dr$
on $r$, or the radial distance from the planet. Spatial distribution
of $dT/dr$ has been previously derived from the results of 
direct numerical simulations of the disk-planet interaction 
(Bate \etal 2003; D'Angelo \& Lubow 2008). At the same time 
there have been few analytical studies of the spatial behavior 
of the torque density. Linear theory of density 
wave excitation has predominantly concentrated, with few 
exceptions (Korycansky \& Pollack 1993, hereafter KP93; 
Goodman \& Rafikov 2001; 
Muto \& Inutsuka 2009), on the behavior of fluid 
variables in Fourier rather than physical space.  

Nevertheless, some asymptotic results regarding the behavior of 
$dT/dr$ have been known since the pioneering study of the
disk-satellite interaction by Goldreich \& Tremaine (1980; hereafter 
GT80). According to this work, far from the perturber,
at radial separations from its orbit $|r-r_p|$ ($r_p$ is the 
semi-major axis of the circular orbit of the planet) exceeding
the disk scale height $h$, the torque density\footnote{We assume
that GT80 calculation refers to $dT/dr$ and not to $dT/dr|_d$ 
since no explicit dissipation mechanism was mentioned in their 
work.} is given by
\ba
\frac{dT}{dr} & \to & \mbox{sgn}(r-r_p)C_{\rm GT80}
\frac{(GM_p)^2\Sigma_0}{\Omega^2|r-r_p|^4},
\nonumber\\
C_{\rm GT80} &=&
\frac{32}{81}\left[2K_0\left(\frac{2}{3}\right)+
K_1\left(\frac{2}{3}\right)\right]^2\approx 2.50783. 
\label{eq:dTdx_GT}
\ea
Here $M_p$ is the planetary mass, $\Sigma_0$ is the disk surface 
density assumed to be uniform on scales $\sim |r-r_p|$,
$\Omega$ is the angular frequency of the disk at $r_p$, and 
$K_n$ is the modified Bessel function of order $n$.

To obtain this expression GT80 have computed the amplitudes of
the torque produced by the individual azimuthal Fourier harmonics of the 
planetary potential and then assigned their action in physical
space to the positions of the corresponding Lindblad resonances. 
With this prescription the one-sided (i.e. considering only $r>r_p$
or $r<r_p$) torque density maintains the same sign independent
of the radial separation from the planetary orbit. Torque density 
behavior 
similar to that given by equation (\ref{eq:dTdx_GT}) was also 
obtained by Lin \& Papaloizou (1979) using impulse approximation.

The torque density prescription (\ref{eq:dTdx_GT}) has been 
extensively used in studies of gap opening
by planets (Lin \& Papaloizou 1986; Trilling \etal 1998; 
Bryden \etal 1999; Armitage \etal 2002; Varnier\'e \etal 2004; 
Crida \etal 2006) and orbital evolution of the SMBH binaries 
surrounded by gaseous disks (Gould \& Rix 2000; Armitage \& 
Natarajan 2002; Lodato \etal 2009; Chang \etal 2010; Alexander \etal 2011). 
Modifications of the prescription (\ref{eq:dTdx_GT}) have 
sometimes been adopted e.g. with the proportionality coefficient 
different from $C_{\rm GT80}$ (e.g. Papaloizou \& Lin 1984; 
Armitage \& Natarajan 2002). However, the key features of the equation 
(\ref{eq:dTdx_GT}) namely the $|r-r_p|^{-4}$ dependence 
and the {\it positive} sign of the asymptotic torque density for
$r>r_p$ (and the negative sign for $r<r_p$) have rarely been 
questioned (cf. Crida \etal 2006).

Recently Dong \etal (2011) investigated the spatial behavior
of the torque density using high-fidelity numerical simulations
of the disk-planet interaction in the two-dimensional shearing 
sheet geometry. Quite unexpectedly, they found that the one-sided 
torque density does not maintain a constant sign but rather 
changes from positive to negative (for $r>r_p$) at 
$|r-r_p|\approx 3.2 h$, in disagreement with the
conclusions of GT80. Dong \etal (2011) have shown this result 
to be quite robust and independent of the numerical 
issues (e.g. resolution of the simulations, simulation box 
size, etc.).

In this work we demonstrate that some of the linear results of 
GT80 need to be revised at the qualitative level even 
within the framework of the linear theory. In 
particular, we show that the torque density does indeed {\it 
change sign} and becomes negative for $r>r_p$ (positive 
for $r<r_p$) beyond several scale heights in radial separation 
$|r-r_p|$ from the planet, in full agreement with the numerical 
results of Dong \etal (2011). This has important 
implications for understanding the issues of 
the gap and cavity opening by planets in protoplanetary disks 
and SMBH binaries in circumbinary disks correspondingly. 

Our paper is structured as follows. In \S \ref{sect:eqs} we
summarize the linear equations for the perturbed fluid 
variables, which are subsequently solved numerically in \S 
\ref{sect:harmonics}. We demonstrate good agreement of the 
results with the analytical calculations based on the global 
Airy representation developed in \S \ref{subsect:Airy_rep}. 
We compute the spatial behavior of the torque density 
and the angular momentum flux in \S \ref{sect:torque} and 
\S \ref{sect:flux} correspondingly . The main 
result of this paper --- confirmation of the negative 
torque density phenomenon --- is described in \S 
\ref{subsect:neg_torque}. 
We discuss the origin of the negative torque density, 
performance of the Airy representation, and compare our results 
with existing studies in \S \ref{sect:disc}. Astrophysical 
implications of our results are outlined in \S 
\ref{sect:implications}.


\section{Basic equations.}  
\label{sect:eqs}


We study tidal coupling of a planet with a uniform disk 
in the shearing sheet geometry (Goldreich \& Lynden-Bell 
1965), which allows us to 
neglect geometric curvature effects while preserving the
main physical ingredients of the system. We
also neglect the vertical dimension and assume the disk to be
two-dimensional. In this approximation one works in a Cartesian
coordinate system with $x=r-r_p$ and $y=r_p(\varphi-\varphi_p)$ 
playing the role of radial and azimuthal coordinates 
correspondingly. 

The dynamics of fluid is then governed by the equations of
motion and continuity (Narayan \etal 1987; hereafter NGG):
\ba
&& \frac{\partial {\bf v}}{\partial t}+
\left({\bf v}\cdot\nabla\right){\bf v}+ 
2{\bf \Omega}\times {\bf v} +4A\Omega z{\bf e}_x
=-\frac{\nabla P}{\Sigma}-\nabla\Phi,
\label{eq:motion}\\
&& \frac{\partial \Sigma}{\partial t}+\nabla\cdot
\left({\bf v}\Sigma\right)=0,
\label{eq:cont}
\ea
where ${\bf v}=(v_x,v_y)$ is the fluid velocity with components in
the $x$ and $y$ directions correspondingly, 
$P$ is gas pressure, $\Phi$ is the planetary potential, 
and $A\equiv (r/2)(d\Omega/dr)$ is the shear rate. When planet is not 
present and the disk is homogeneous, $\Sigma=\Sigma_0$, these 
equations have a solution in the form of a simple linear shear: 
$v_{x,0}=0,~v_{y,0}=2Ax$. 

When a planet of mass $M_p$ is present (at $x=y=0$) 
a stationary (i.e. $\partial/\partial t=0$) pattern of the 
density and velocity perturbations gets established in the disk.
We can then analyze equations (\ref{eq:motion})-(\ref{eq:cont}) in a 
standard way, see GT80 and NGG for details. First,
we assume $\Sigma=\Sigma_0+\Sigma_1$, $v_x=v_{x,1}$, $v_y=2Ax+v_{y,1}$
and linearize equations for small perturbations $\Sigma_1$,
$v_{x,1}$ and $v_{y,1}$. We then represent all perturbed fluid 
variables via Fourier integrals as 
$\left\{\Sigma_1,v_{x,1},v_{y,1},\Phi\right\}=
(2\pi)^{-1}\int_{-\infty}^\infty dk_y\exp\left(ik_y y\right)
\left\{\delta\Sigma,u,v,\phi\right\}$, which 
takes care of the $y$-dependence leaving us with a set
of equations in $x$ coordinate only.

NGG have shown that the resulting system can be reduced to a 
single differential equation for the azimuthal velocity 
perturbation $v$ only:
\ba
\frac{\partial^2 v}{\partial x^2}+v\left(\frac{4A^2k_y^2}{c^2}x^2
-\frac{\kappa^2+k_y^2c^2}{c^2}\right)=
-\frac{2Ak_y^2}{c^2}x\phi-\frac{2B}{c^2}\frac{\partial \phi}
{\partial x},
\label{eq:v}
\ea
while the perturbed radial velocity $u$ and the surface density 
perturbation $\delta\Sigma$ are directly expressed in terms of 
$v$ via the following relations:
\ba
&& u=-\frac{i}{k_y^2c^2+4B^2}\left(k_y c^2\frac{\partial v}
{\partial x}+4 A B k_y x v + 2Bk_y\phi\right),
\label{eq:u}\\
&& \delta\Sigma=\frac{\Sigma_0}{k_y^2c^2+4B^2}\left(2B\frac{\partial v}
{\partial x}-2 A k_y^2 x v - k_y^2\phi\right).
\label{eq:dS}
\ea
Here $c$ is the sound speed, $B=\Omega+A$ is the Oort's constant, 
$\kappa=(4B\Omega)^{1/2}$ is the epicyclic 
frequency and $\phi$ is the Fourier component of the planetary 
potential $\Phi$ given in a point-mass case by 
\ba
\phi(k_y,x)=-\frac{GM_p}{\pi}K_0\left(|k_yx|\right),~~~~~~~
\frac{\partial \phi}{\partial x}=\sgn(x)\frac{GM_p}{\pi}k_y 
K_1\left(|k_yx|\right).
\label{eq:pot}
\ea
In a Keplerian disk $A=-(3/4)\Omega$, $B=\Omega/4$ and 
$\kappa=\Omega$.

Introducing a new dimensionless radial coordinate 
\ba
z(x,k_y)\equiv x(4|A|k_y/c)^{1/2}
=\frac{x}{h}\left(\frac{4|A|}{\Omega}k_yh\right)^{1/2},~~~
x(z,k_y)=zh\left(\frac{4|A|}{\Omega}k_yh\right)^{-1/2},
\label{eq:var_trans}
\ea 
where $h=c/\Omega$ is the scale height,
we rewrite equation (\ref{eq:v}) in
a standard parabolic cylinder equation form 
\ba
&& \frac{\partial^2 v}{\partial z^2}+v\left(\frac{z^2}{4}-a\right)=
R(k_y,x),
\label{eq:parab}\\
&& R(k_y,x)\equiv
\frac{k_y x}{2c}\phi-\frac{B}{2|A|k_y c}\frac{\partial \phi}
{\partial x}=-\frac{GM_p}{2\pi c}\left[
k_yxK_0\left(|k_yx|\right)+\sgn(x)\frac{B}{|A|}
K_1\left(|k_yx|\right)\right].
\label{eq:R}
\ea
Here
\ba
a=\frac{\kappa^2+k_y^2c^2}{4|A|k_yc}=\frac{\kappa^2}{4|A|\Omega}
\frac{1+(k_yh)^2\left(\Omega/\kappa\right)^2}{k_yh}
=\frac{1+(k_yh)^2}{3(k_yh)},
\label{eq:C2}
\ea
where the last equality holds for a Keplerian disk.
Clearly, $a\gg 1$ both when $k_yh\ll 1$ and $k_yh\gg 1$, i.e.
for the modes excited at Lindblad 
resonances lying either very close to the planet, at $|x|\lesssim h$,
or far from the planet, at $|x|\gtrsim h$. Density wave harmonics
with $k_yh\sim 1$, which carry most of the angular momentum flux 
in a disk of uniform surface density and are excited at $|x|\sim h$ 
have $a\sim 1$.

Equation (\ref{eq:parab}) must be solved with the boundary condition 
in the form of the outgoing wave for $x\to \pm\infty$.


\section{Solutions of fluid equations.}  
\label{sect:harmonics}


Solution of equation (\ref{eq:parab}) can be directly expressed 
in terms of the parabolic cylinder functions, but they are not
easy to analyze. Thus, instead of working directly with the 
parabolic cylinder functions we solved equation (\ref{eq:parab})
numerically as outlined below in \S \ref{subsect:numerics}.
Also, as we describe in \S \ref{subsect:Airy_rep} 
and Appendix \ref{sect:Airy}, whenever $a\gtrsim 1$ one 
can globally approximate the solutions of equation (\ref{eq:parab}) via 
the combinations of Airy functions of special arguments. 
We call this approximation the 
{\it Airy representation} and it forms the basis for our 
subsequent analysis of the disk-planet coupling.


\subsection{Numerical procedure.}  
\label{subsect:numerics}

Direct numerical integration of the equation (\ref{eq:v}) 
is based on the procedure developed in KP93, where a 
shooting method is implemented by matching the numerical 
solution from the origin to the WKB outgoing waves. 

As shown in NGG, the exact homogeneous solution to
equation (\ref{eq:v}), i.e. the parabolic 
cylinder function, can be well approximated by the WKB outgoing 
waves
\ba
v(x\to\pm\infty)\sim\sqrt{\frac{2}{3k_{y}x}}e^{\pm i3k_{y}x^{2}/4h}
\label{eq:v_wkb}
\ea
(in a Keplerian disk) when $x/h\gg (8/9)(1+(k_{y}h)^{2})(k_{y}h)^{-2}$. 
Additionally, for this approximation to accurately represent the 
inhomogeneous solution of (\ref{eq:v}) one requires $|xk_y|\gg1$, 
so $\phi\ll1$ and $ \partial\phi/\partial x\ll1$. Both conditions 
together imply that
$x/h \gg \min\{1/(k_yh),1\}$, which is the criterion we use to 
match our numerical
solutions to the outgoing waves.

Following KP93, to obtain solutions satisfying boundary conditions 
we start  by shooting two arbitrary linearly 
independent homogeneous 
solutions $v_{1}^{h}$ and $v_{2}^{h}$ and one arbitrary 
inhomogeneous solution $v^{i}$ from the origin to the 
boundaries at $\pm x_B$ with $x_B/h \gg \min\{1/(k_yh),1\}$. 
Then we can write the desired solution as a linear combination 
of these solutions
\ba
v=v^{i}+a_{1}v_{1}^{h}+a_{2}v_{2}^{h},
\end{eqnarray}
where $a_{1}$ and $a_{2}$ are two complex constants to be 
determined by the boundary conditions following from the 
equation (\ref{eq:v_wkb}): 
\begin{eqnarray}
&&\frac{\partial }{\partial x}
\left\{v^{i}(-x_B) + a_{1}v_{1}^{h}(-x_B)+a_{2}v_{2}^{h}(-x_B)\right\}
\nonumber\\
&&=\left( i\frac{3}{2}k_{y} x_{B} +\frac{1}{6k_{y}  x_{B}} \right)
\left\{v^{i}(-x_B)+a_{1}v_{1}^{h}(-x_B)+a_{2}v_{2}^{h}(-x_B) \right\}
,\\
&&\frac{\partial }{\partial x}
\left\{v^{i}(x_B) + a_{1}v_{1}^{h}(x_B)+a_{2}v_{2}^{h}(x_B)\right\}
\nonumber\\
&&=\left( i\frac{3}{2}k_{y} x_{B} - \frac{1}{6k_{y}x_{B}} \right)
\left\{v^{i}(x_B)+a_{1}v_{1}^{h}(x_B)+a_{2}v_{2}^{h}(x_B)\right\}.
\label{eq:boundary}
\ea

This method is quite efficient since we only 
have to integrate three solutions and invert a $2\times 2$ matrix 
to find an exact solution for a
given $k_y$. Moreover, we have numerically proven that $x_B/h$ does 
not need to be too large compared to $\min\{1/(k_yh),1\}$ to  
give the right answer with good accuracy.

In the subsequent numerical calculations we use a 4-th order 
Runge-Kutta integrator with spatial resolution of $h/200$ for 
320 uniformly log-spaced values of $k_yh$ between 0.01 and 15, 
and potential softening length of $10^{-4}h$.

\begin{figure}
\plotone{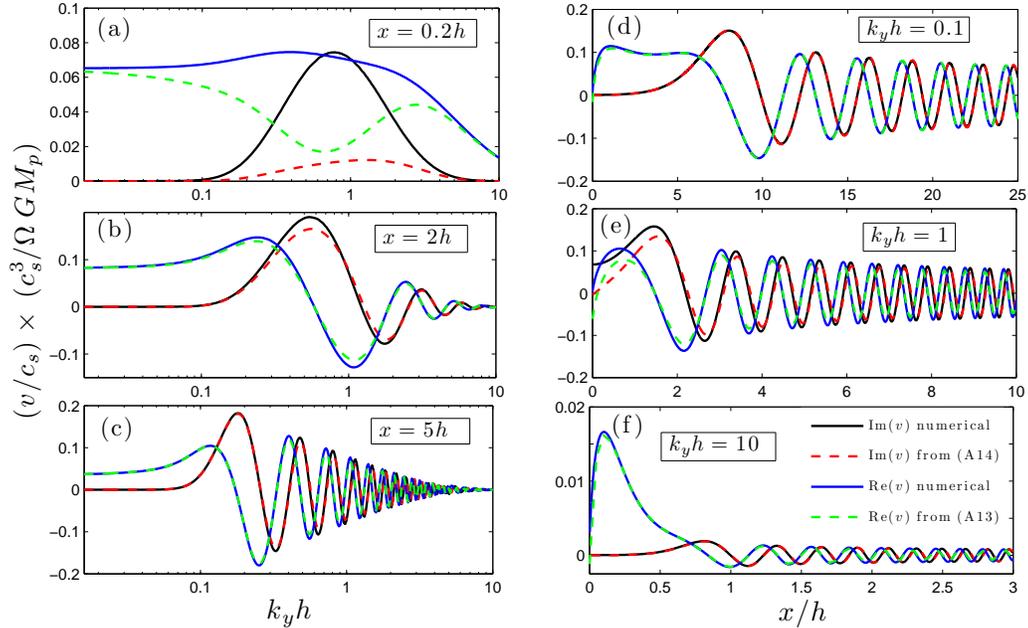}
\caption{Behavior of the individual Fourier harmonics of the
azimuthal velocity perturbation $v$ as a function of both $k_y$ 
(left panels) and $x$ (right panels). Respective values of $x$ 
or $k_yh$ are shown in each panel. We plot $\Re(v)$ and 
$\Im(v)$ obtained both by the direct numerical integration 
of the equation (\ref{eq:v}) (solid curves) and using the Airy
representation (dashed curves). 
\label{fig:v_k}}
\end{figure}


\subsection{Analytical Airy representation.}  
\label{subsect:Airy_rep}

In Appendix \ref{sect:Airy} we show that the solution of equation
(\ref{eq:parab}) with the boundary condition in the form of 
outgoing waves can be expressed in the limit $a\gg 1$ via the
Airy functions, which have some algebraic function of $x$ as 
their argument, see equations (\ref{eq:Esols})-(\ref{eq:tau}). 
There we also derive the expressions (\ref{eq:Rev_full}) and 
(\ref{eq:Imv_full}) for the real and imaginary parts of $v$,
which contain a number of exponentially small terms.  
Neglecting these terms, which is valid when $a\gg 1$, i.e. when 
either $k_yh\gg 1$ or $k_y h\ll 1$ one finds that the solution 
for $v$ reduces to
\ba
&& \Re(v)\sim -\frac{\pi g(z)}{a^{1/2}}\left[
\Ai(-t(z))\int\limits^{\infty}_z g(s)\Bi(-t(s))R(s)ds+
\Bi(-t(z))\int\limits_0^z g(s)\Ai(-t(s))R(s)ds\right],
\label{eq:Rev}\\
&& \Im(v)\sim -\frac{\pi g(z)}{a^{1/2}}I_+
\Ai(-t(z)),
\label{eq:Imv}\\
&& I_+ = \int\limits_0^{\infty} g(s)
\left[\Ai(-t(s))-\frac{e^{-\pi a}}{2}
\Bi(-t(s))\right]R(s)ds.
\label{eq:I-def}
\ea
Here $\Ai$ and $\Bi$ are Airy functions, while functions $g$ and $t$
are defined by equations (\ref{eq:defs1})-(\ref{eq:tau}).

We also demonstrate in Appendix \ref{sect:Airy} that whenever 
$k_yh\lesssim 1$ and $a\gtrsim 1$ the factor $I_+$ defined 
by equation (\ref{eq:I-def}) can be approximated for a 
Keplerian disk as
\ba
I_+=R(2\sqrt{a}h(3k_yh)^{-1/2})=-\frac{GM_p}{2\pi c}
\left[wK_0\left(|w|\right)+\frac{1}{3}
K_1\left(|w|\right)\right],~~~~~w=\frac{2}{3}\sqrt{1+(k_yh)^2}.
\label{eq:I}
\ea
Equations (\ref{eq:Rev})-(\ref{eq:I}) provide the foundation for
our subsequent linear analytical study of the planet-generated density 
waves in \S\S \ref{sect:torque}, \ref{sect:flux}.


\subsection{Results.}  
\label{subsect:results}

In Figure \ref{fig:v_k} we show the behavior of the Fourier 
harmonics of the
azimuthal velocity perturbation $v$ as a function of $x$ for 
a fixed $k_y$ and also as a function of $k_y$ for a fixed $x$.
We display the behavior of $\Re(v)$ and $\Im(v)$ obtained by
numerically solving equation (\ref{eq:parab}), and also the analytical 
Airy representation of the same variable given by equations 
(\ref{eq:Rev_full})-(\ref{eq:Imv_full}). 

\begin{figure}
\plotone{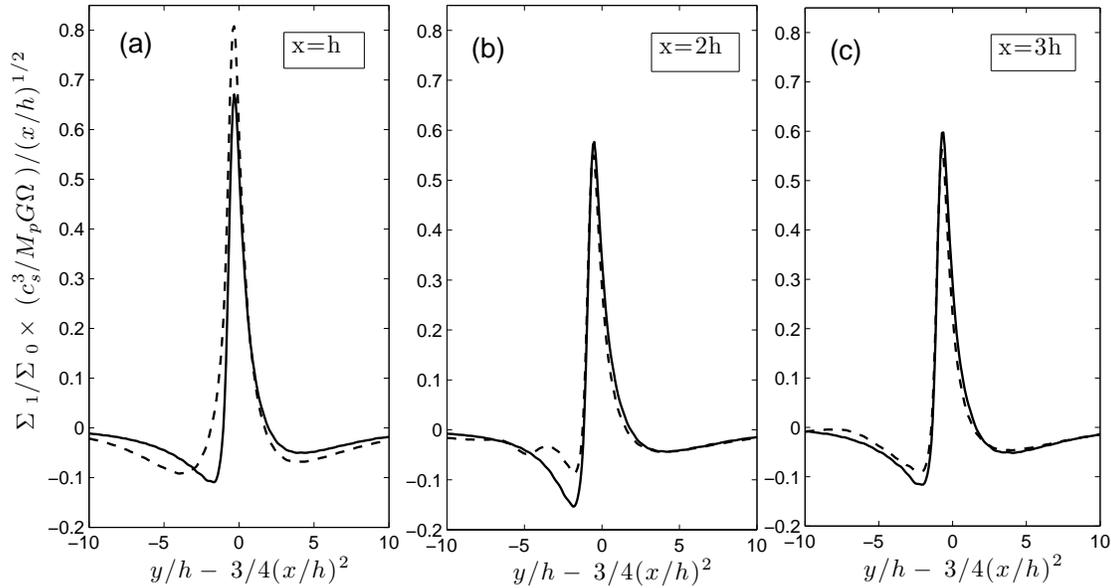}
\caption{Azimuthal cuts through the density wake in coordinate 
space showing the surface density perturbation $\Sigma_1$ at 
different radial separations from the planet ($x=h,2h,3h$
as labeled), normalized by $x^{1/2}$. Solid curves show numerical 
results while the dashed curves correspond to $\Sigma_1$ profiles
obtained using Airy representation. 
An azimuthal shift by $(3/4)x^2$ is applied in each panel to 
facilitate comparison.
\label{fig:dsigma}}
\end{figure}

For a fixed $k_y$ satisfying the conditions $k_yh\lesssim 1$ or 
$k_yh\gtrsim 1$ (i.e. for $a\gtrsim 1$) the behavior of $v$ 
is reproduced by the Airy representation very well --- in most 
cases the analytical prediction is hardly discernible from 
the numerical calculation, see Figures \ref{fig:v_k}d,f. 
But as Figure \ref{fig:v_k}e shows the Airy 
representation actually works quite well even for $k_yh\sim 1$
when formally it should not be applicable: analytical prediction 
falls essentially on top of the numerical results for both 
$\Re(v)$ and $\Im(v)$ as long as $x\gtrsim 5h$; significant 
discrepancy between the two is noticeable only for $x\lesssim 2h$.

Results for fixed $x$ tend to confirm this trend. Already at 
moderately large separations from the planet, e.g. $x=5h$, 
analytical predictions (\ref{eq:Rev_full})-(\ref{eq:Imv_full}) 
reproduce the numerical results for all
values of $k_y$ (including $k_yh\sim 1$) extremely well, see 
Figure \ref{fig:v_k}c. Even at $x=2h$ the agreement is still 
pretty good, see Figure \ref{fig:v_k}b, even though at 
this value of $x$ the modes with $k_yh\sim 1$ play the 
dominant role. It is only very close to the planet, e.g. at $x=0.2h$
as shown in Figure \ref{fig:v_k}a, that the discrepancy between the Airy 
representation and the direct numerical calculation becomes 
significant, but again only for $k_yh\sim 1$ (analytical formulae
agree with the numerical results in the limits $k_yh\ll 1$ and 
$k_yh\gg 1$ even at $x=0.2h$).

In Figure \ref{fig:dsigma} we plot the behavior of the surface 
density perturbation $\Sigma_1(x,y)$ in physical space. We 
compute $\Sigma_1$ fully numerically by calculating its Fourier 
harmonics using equation (\ref{eq:dS}) with numerically determined 
$v$ and then taking the inverse Fourier integral. We also calculate 
$\Sigma_1$ semi-analytically by repeating the same procedure 
with $\Re(v)$ and $\Im(v)$ given by equations 
(\ref{eq:Rev_full})-(\ref{eq:Imv_full}). One can see that 
the two ways of computing $\Sigma_1(x,y)$ agree with each 
other quite well. They do show some discrepancy at small $x\approx h$,
which is expected given the dominant role of the modes with
$k_yh\sim 1$ in shaping the harmonic content of $\Sigma_1(x,y)$
close to the planet. However, further out from the planet 
the agreement between the Airy representation and the numerical 
calculation improves, see Figure \ref{fig:dsigma}c. Azimuthal density
wave profiles at different separations from the planet have previously
been computed by Goodman \& Rafikov (2001) with the same method as
we employ here, and by Dong \etal (2011) using direct numerical 
hydrodynamical simulations of tidal coupling between the low mass
planet and the disk. Their results are in agreement with ours. 

These comparisons clearly illustrate the robustness of the 
analytical Airy representation for the azimuthal velocity 
perturbation $v$ in certain limits. They also show that 
this representation can be used for accurate 
calculation of the behavior of the fluid variables in physical 
space at large $x/h$. All this gives us confidence in 
the subsequent use of this analytical approximation for the 
calculation of the asymptotic behavior of the torque density
in \S \ref{sect:torque} and the angular momentum flux in
\S \ref{sect:flux}.


\section{Torque density.}  
\label{sect:torque}


We now proceed to study the spatial 
behavior of the torque density $dT/dx$ --- the amount of 
torque that is exerted by a planet on the disk per unit 
radial distance $x$. Analytical calculation
of this important physical quantity has been first carried
out by GT80 but only in the asymptotic regime $|x|\gtrsim h$.
Besides, as we show below, this calculation was incorrect.  
Also, a number of numerical studies have derived $dT/dx$
for arbitrary $x$ as a by-product of their simulations
(Bate \etal 2003; D'Angelo \& Lubow 2008).
Our current semi-analytical calculation is intended to 
provide the full description of $dT/dx$ in the linear regime 
for any $x$.  

The torque density is defined as
\ba
\frac{dT}{dx}=-\int\limits_{-\infty}^\infty
dy\Sigma_1\frac{\partial\Phi}{\partial y}.
\label{eq:T_H}
\ea
Note that compared to $dT/dr$ this definition lacks an extra 
factor $r_p$ and strictly speaking represents the momentum 
density (even though we will still call it torque density 
in this work). This is because $r_p$ is not a well defined 
variable in the shearing sheet setup.
 
\begin{figure}
\plotone{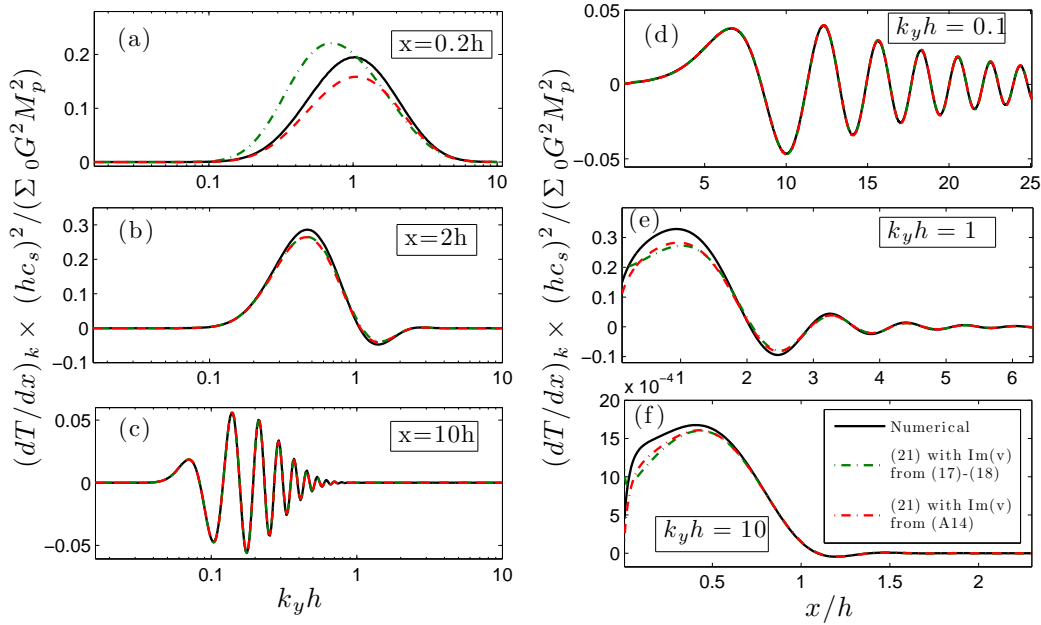}
\caption{Behavior of the individual Fourier harmonics of the
torque density $(dT/dx)_k$ as a function of both $k_y$ 
(left panels) and $x$ (right panels), labeled correspondingly. 
Solid curves show
numerical results while the dashed and dot-dashed curves represent 
$(dT/dx)_k$ obtained using different levels of the Airy 
representation (see text for details).
\label{fig:dTdx_k}}
\end{figure}

Fourier decomposing $\Sigma_1$ and $\Phi$ in azimuthal
direction and manipulating the resultant expression we
arrive at the following alternative expression:
\ba
&&\frac{dT}{dx}=\int\limits_0^\infty \left(\frac{dT}{dx}\right)_k dk_y,
\nonumber\\
&&\left(\frac{dT}{dx}\right)_k=-4\pi k_y\phi\Im(\delta\Sigma)
=-\frac{8\pi\Sigma_0}{k_y^2c^2+4B^2}
k_y\phi\left[B\frac{\partial}{\partial x}\Im(v)-
Ak_y^2x\Im(v)\right].
\label{eq:T_Hk}
\ea
The last equality was derived using equation (\ref{eq:dS}).

Our numerical solutions for the azimuthal velocity perturbation $v$
obtained in the previous section by numerically integrating the 
linear fluid equation (\ref{eq:parab}) allow us to compute the individual 
Fourier harmonics $(dT/dx)_k$ using equation (\ref{eq:T_Hk}).
In Figure \ref{fig:dTdx_k} we show the behavior of $(dT/dx)_k$ as
a function of both $k_y$ and $x$. We also display $(dT/dx)_k$
computed in the framework of the Airy representation using two 
levels of accuracy. Dot-dashed curve is derived from equation 
(\ref{eq:T_Hk}) using $v$ given by equation (\ref{eq:Imv}), 
which neglects the exponentially small terms, with $I_+$ 
given by equation (\ref{eq:I-def}). The $(dT/dx)_k$ shown 
by the dashed line uses $v$ given by equation (\ref{eq:Imv_full}) in which 
all subdominant terms have been retained. 

Again, we see very good agreement 
between the theoretical Airy representation and the numerical results
in the limits of both $k_y h\gtrsim 1$ and  $k_y h\lesssim 1$ for all
$x$, as well as in the limit of $x\gtrsim h$ for any $k_y h$. 
This is 
in full agreement with the results for $v$ shown in Figure 
\ref{fig:v_k} since $(dT/dx)_k$ is directly derived from $v$
using (\ref{eq:T_Hk}). One can also see that the use of equation 
(\ref{eq:Imv_full}) instead of (\ref{eq:Imv}) to represent $v$
gives a more accurate approximation of the numerical results. For that
reason we generally recommend using the more accurate equation 
(\ref{eq:Imv_full}) in semi-analytical calculations 
involving the Airy representation.

Next we explore the behavior of the full torque density $dT/dx$, 
integrated over all $k_y$ according to the equation (\ref{eq:T_Hk}). 
In Figure \ref{fig:dTdx} we present the result of such calculation 
using numerically derived $v$. We also plot the analytical $dT/dx$ 
computed using Airy representation of $v$ at two levels of accuracy. 
First, simplified calculation (dot-dashed line) uses $\Im(v)$ given 
by equation (\ref{eq:Imv}) with $I_+$ computed according to the 
approximation (\ref{eq:I-def}). Not surprisingly this method of 
computing $dT/dx$ does not fare well at $x\sim h$ and it shows 
significant discrepancy with the numerical solution as $x/h\to 0$.
Second, more accurate version of the Airy representation 
(dashed line) relies
on equation (\ref{eq:Imv_full}) retaining the exponentially small 
terms (which, however, become important as $k_yh\sim 1$) to 
describe the behavior of $\Im(v)$. This calculation yields 
better agreement with the numerical result as shown in Figure 
\ref{fig:dTdx} even though there is still significant discrepancy 
(at the level of $10-20\%$) as $x\sim h$. At the same time, as 
expected based on the results of \S \ref{sect:harmonics} both 
versions of the Airy representation work quite well at
large separations from the planet, at $x\gtrsim 2h$.

We also plot in the Figure \ref{fig:dTdx} $dT/dx$ 
derived from the direct numerical simulations of the
disk-planet interaction (Dong \etal 2011). Clearly, the 
agreement between the linear theory results and the outcome of 
direct simulations is very good. 

\begin{figure}
\plotone{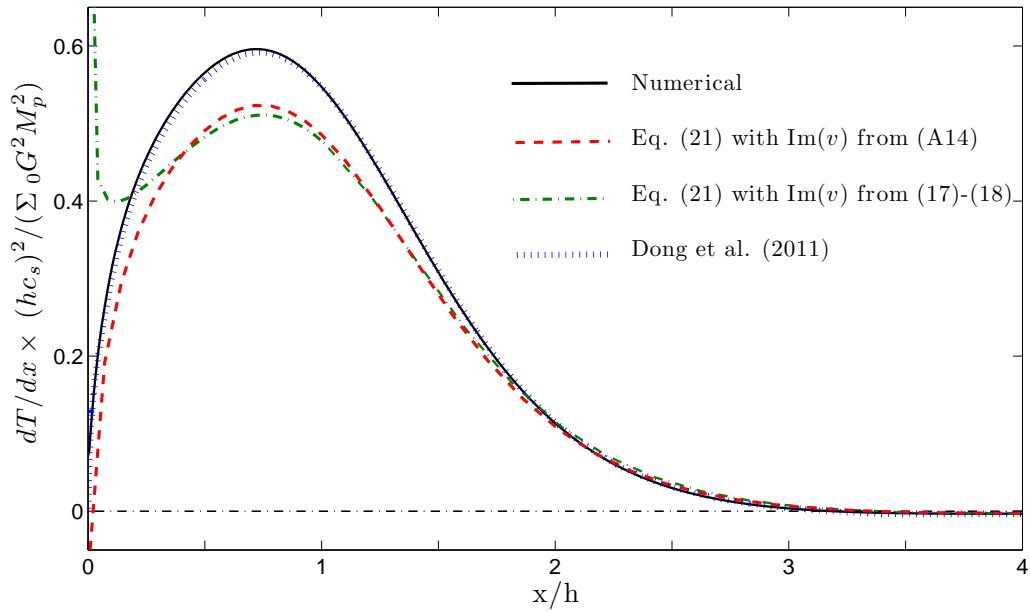}
\caption{Full torque density $dT/dx$ (integrated over all $k_y$) 
as a function of $x$ in the vicinity of the planet. Numerical 
linear results are given by solid line. Results obtained with the Airy 
representation (equation (\ref{eq:T_Hk}) integrated using $\Im(v)$ 
given by equation (\ref{eq:Imv_full})) are shown by dashed line, and 
those using $\Im(v)$ from equations (\ref{eq:Imv})-(\ref{eq:I-def}) 
are shown by dot-dashed line. Dotted curve shows $dT/dx$ extracted 
from one of the numerical simulations of Dong \etal (2011). 
\label{fig:dTdx}}
\end{figure}


\subsection{Negative torque density phenomenon.}  
\label{subsect:neg_torque}

Different methods of calculating $dT/dx$ shown in Figure 
\ref{fig:dTdx} universally exhibit the phenomenon of the negative 
torque density first described in Dong \etal (2011). 
Using direct numerical simulations these authors have shown that 
in a uniform disk $dT/dx$ changes sign at the finite separation 
from the planet in apparent disagreement with the predictions of 
GT80. 

To illustrate this behavior in more detail in 
Figure \ref{fig:neg_torque}a we show a zoomed-in view of $dT/dx$ 
at large separations $x>3h$. There we plot results of our numerical
calculation, analytical calculation using Airy representation with 
$\Im(v)$ from equation (\ref{eq:Imv_full}), and $dT/dx$ derived 
from the direct numerical simulations of
Dong \etal (2001). All three methods of computing $dT/dx$ 
agree on the distance $x_-\approx 3.2h$ where $dT/dx$ changes 
sign. Linear theory results (both numerical and analytical) show 
good agreement with each other beyond this point while the $dT/dx$ 
derived from the direct hydrodynamical simulations exhibits some 
features related to numerical issues (mainly determined by the 
finite extent of the simulation domain in $x$-direction, see 
Dong \etal 2011).

We also show on this plot the asymptotic behavior of $dT/dx$
according to GT80 (thin solid line), see equation (\ref{eq:dTdx_GT}). 
It clearly disagrees 
with the linear results obtained in this work and with the
results of the simulations by Dong \etal (2011)
both in amplitude and the sign of the effect at $x\gtrsim 1$. 
We provide explanation for this discrepancy in \S 
\ref{sect:origin}.

In Appendix \ref{sect:T_H} we show using Airy representation 
of $v$ that the asymptotic behavior of the torque density in a 
Keplerian disk valid in the limit $x/h\to\infty$ is given by 
\ba
\left(\frac{dT}{dx}\right)^{as} & \to & \mbox{sgn}(x) 
C\frac{(GM_p)^2\Sigma_0}{\Omega^2}\frac{1}{x^4},\nonumber
\\
C & = & \frac{16}{81}
\left[2K_1\left(\frac{2}{3}\right)-5K_0\left(\frac{2}{3}\right)\right]
\left[2K_0\left(\frac{2}{3}\right)+K_1\left(\frac{2}{3}\right)\right]
\approx -0.613096.
\label{eq:dTdx_full}
\ea
This expression replaces equation (\ref{eq:dTdx_GT}), which 
previously described the asymptotic behavior of $dT/dx$ 
according to GT80.

In Figure \ref{fig:neg_torque}b we compare the asymptotic scaling 
given by equation (\ref{eq:dTdx_full}) with the behavior 
of $dT/dx$ at large separations ($x<12 h$) computed using our 
numerical determination of $v$ in equation (\ref{eq:T_Hk}). 
We also show on this plot $dT/dx$ obtained by numerically integrating 
the analytical expression (\ref{eq:T_Hk_withAiry1}) over $k_y$. 
One can see that all three ways of representing the behavior
of $dT/dx$ at large $x$ agree with each other quite well.

\begin{figure}
\plotone{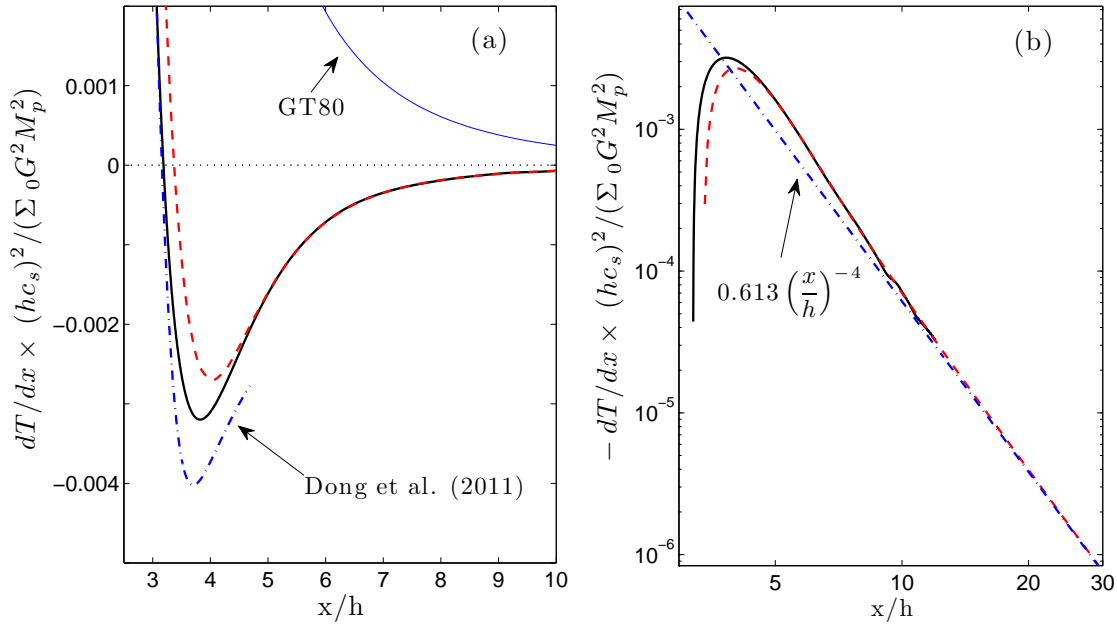}
\caption{Illustration of the negative torque density phenomenon. 
(a) Behavior of $dT/dx$ for $x>3h$ clearly showing change
of sign of the torque density at $x_-\approx 3.2 h$. Solid line 
stands for numerical linear calculation, dashed curve is for Airy
representation with $\Im(v)$ given by equation (\ref{eq:Imv_full}),
and dot-dashed line shows the results of simulations (Dong \etal 
2011). Thin solid line shows the asymptotic prediction of GT80. 
(b) Asymptotic behavior of $dT/dx$ at large $x/h$ (note the negative 
sign). Solid curve is the same as in panel (a), dot-dashed line shows 
asymptotic behavior (\ref{eq:dTdx_full}), and dashed line results 
from integrating equation (\ref{eq:T_Hk_withAiry1}) over $k_y$. 
\label{fig:neg_torque}}
\end{figure}


\section{Angular momentum flux.}  
\label{sect:flux}



To get the full picture of the disk evolution driven by 
planet-generated density waves it is important to properly 
understand not only the wave damping mechanisms but also 
the amount of the angular momentum carried by the waves ---
angular momentum flux $F_H$, and its spatial pattern.

Total angular momentum flux carried by the density wave is
\ba
F_H=\Sigma_0\int\limits_{-\infty}^\infty
dy~ v_{x,1}v_{y,1}
=\int\limits_0^\infty F_{H,k}dk_y,~~~~~
F_{H,k}=4\pi\Sigma_0\left[\Re(v)\Re(u)+\Im(v)\Im(u)\right]
=4\pi\Sigma_0\Re\left(u v^*\right),
\label{eq:F_H}
\ea
where $F_{H,k}$ is the angular momentum flux carried by a 
particular azimuthal mode of the wave, and $*$ denotes complex 
conjugate. Using equation (\ref{eq:u}) and the fact that $\phi$
is purely real we find
\ba
F_{H,k}=-\frac{4\pi\Sigma_0}{k_y^2c^2+4B^2}\left[
k_yc^2\Re\left(iv^*\frac{\partial v}{\partial x}\right)+
2Bk_y\phi\Im(v)\right]
\label{eq:F_Hk}
\ea
Because of the angular momentum conservation the radial divergence
of the angular momentum flux $dF_{H,k}/dx$ must be equal to 
$(dT/dx)_k$. To demonstrate this property we differentiate 
equation (\ref{eq:F_Hk}) with respect to $x$ and then
transform the term proportional to $v^*\partial v/\partial x$
using equation (\ref{eq:v}). In the end we find that 
$d F_{H,k}/d x$ is indeed given by the same expression in 
terms of $\Im(v)$ as $(dT/dx)_k$, see equation (\ref{eq:T_Hk}).

In the important limit $x\to \infty$ we can compute the asymptotic 
value of the angular momentum flux harmonic $F_{H,k}$ analytically. 
For this purpose we use equation (\ref{eq:F_Hk}), in which 
$\phi\to 0$ as $x\to \infty$ (see equation (\ref{eq:R})), so that 
\ba
F_{H,k}(x\to \infty)\to\frac{4\pi\Sigma_0k_yc^2}{k_y^2c^2+4B^2}\left[
\Re(v)\frac{\partial}{\partial x}\Im(v)-
\Im(v)\frac{\partial}{\partial x}\Re(v)\right].
\label{eq:F_Hk_far}
\ea
In the same limit $x,z\to \infty$ equation (\ref{eq:Rev})
reduces to
\ba
\Re(v)\sim -\frac{\pi g(z)}{a^{1/2}}I_+\Bi(-t(z)),
\label{eq:Rev_far}
\ea
where $I_+$ is given by equation (\ref{eq:I-def}). The exponentially
small contributions present in equation (\ref{eq:Rev_full}) have 
been dropped from this expression.

Using these facts we show in Appendix \ref{sect:F_H} that the
Fourier component of the angular momentum flux is given in 
the limit $k_yh\lesssim 1$ and $x/h\gg 1$ by the following 
formula in a Keplerian disk:
\ba
F_{H,k}(x\to \infty)=\frac{4}{3}k_y^2\frac{(GM_p)^2\Sigma_0}{\Omega^2}
\left[2K_0\left(\frac{2}{3}\right)+K_1\left(\frac{2}{3}\right)\right]^2.
\label{eq:fin}
\ea
This expression coincides with the corresponding result of
GT80 obtained in the limit $m\lesssim r/h$ (see their equation (91)).

In Figure \ref{fig:F_H} we present the run of the full angular
momentum flux $F_H(x)$ in real space, as a function of the radial 
separation from the planet. One can see that $F_H$ starts with 
a non-zero value of $0.098$ (in units of 
$(G^2M_p^2\Sigma_0)/(hc^2)$) at $x=0$ and then gradually 
rises until $x$ is about $3$. Beyond that point $F_H(x)$ 
starts decreasing with the distance although the decline 
is rather modest: the peak value of $F_H$ is $0.938$ (in the 
same natural units) while the value of $F_H$ at $x=15h$ is 
about $0.908$. This reduction of $F_H(x)$ with the distance
is a direct consequence of the negative torque density 
phenomenon discussed in \S \ref{subsect:neg_torque}, since 
$dF_H/dx=dT/dx$ as we have shown before. The latter point is 
additionally emphasized in Figure \ref{fig:F_H} where we also plot the 
integrated torque $T(x)\equiv \int_0^x(dT/dx)dx$. One can see 
that $F_H(x)$ and $T(x)$ are identical to each other up to 
the vertical shift by $F_H(0)$.

Previously, using calculation in the Fourier space, 
GT80 have found $F_H(\infty)=0.93$, which is close to the value we 
find. The small difference between these numbers might possibly  
be ascribed to the small $k_y$ range used by GT80 to compute 
$F_H(\infty)$ and should not be taken too seriously.

\begin{figure}
\plotone{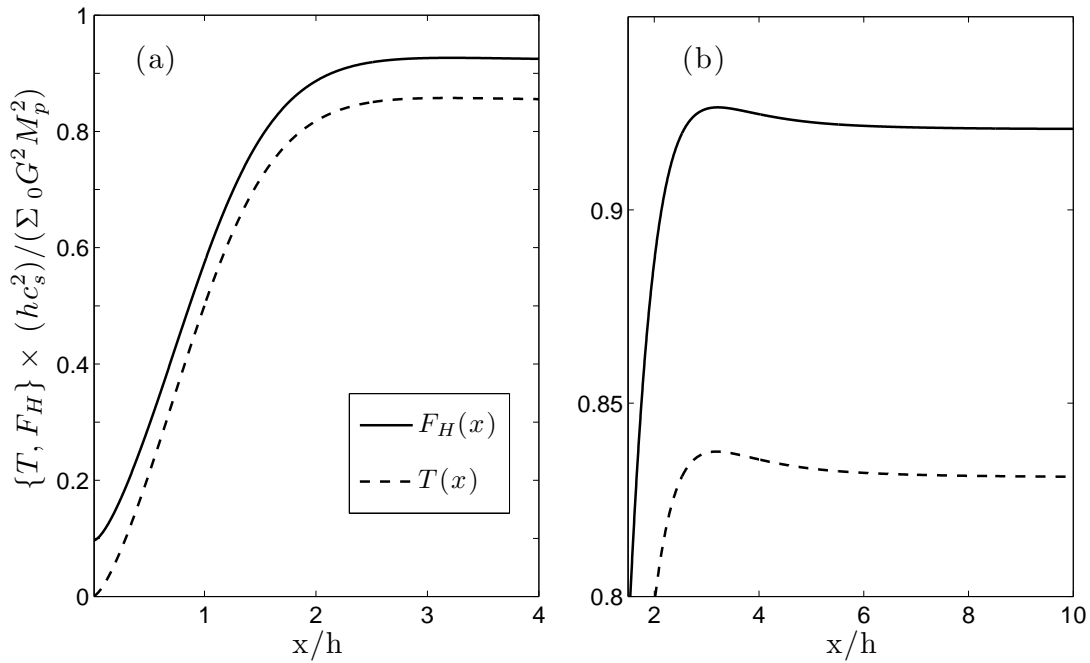}
\caption{(a) Scaling of the angular momentum flux $F_H$ with the 
distance from 
the planet (solid curve). We also plot integrated torque $T(x)$ 
accumulated by the wave at distance $x$ (dashed curve). Note 
that $F_H(x)$ has 
a non-zero value at $x=0$, and that $dT/dx=dF_H/dx$. (b) Close-up 
view at $x>2h$ to illustrate the slight decay of $F_H(x)$ and 
$T(x)$ at large $x$ as a result of the negative torque density
phenomenon (see \S \ref{subsect:neg_torque}).
\label{fig:F_H}}
\end{figure}


\section{Discussion.}
\label{sect:disc}


Most of the existing linear calculations of the disk-satellite
coupling have been limited to the Fourier space  
(Ward 1986; Artymowicz 1993; Tanaka \etal 2002). 
Notable exceptions include Meyer-Vernet 
\& Sicardy (1987) who discussed the radial structure of the 
perturbed velocity eigenfunctions for a fixed azimuthal 
wavenumber under a variety of assumptions about the physics 
operating in the disk. Later on KP93 have done similar 
calculation for a disk with non-zero pressure (the same setup 
as used in our current work) and computed the run of $F_{H,k}$ 
as a function of $x$. But these studies have only looked at 
the spatial behavior of {\it individual} modes driven by 
the planetary potential.

More recently Goodman \& Rafikov (2001) have integrated the 
contributions of separate modes over the azimuthal 
wavenumber to obtain the full spatial distribution of 
the perturbed surface density. Our current 
work extends the study Goodman \& Rafikov (2001) by 
computing the spatial behavior of other important fluid
variables such as the torque density and the angular momentum 
flux.

The key result of our present study is the confirmation of the 
negative torque density phenomenon (discovered in the numerical 
study of Dong \etal (2011)) in the framework of the
linear theory of disk-satellite interaction, and the determination
of the asymptotic behavior of $dT/dx$ far from the planet.
The discrepancy between our result (\ref{eq:dTdx_full}) and 
the conventional expression (\ref{eq:dTdx_GT}) not just in 
amplitude of the effect but also in its sign quite naturally
demands an explanation, which we provide below.


\subsection{Origin of the negative torque density.}
\label{sect:origin}

All calculations of the torque and angular momentum flux 
excited by the planet in GT80 have been carried out in 
Fourier space for individual harmonics of the planetary potential. 
These results fully agree with our calculations in 
\S\S \ref{sect:torque} \& \ref{sect:flux}, in particular 
we successfully reproduce their expression for $F_{H,k}$
in \S \ref{sect:flux}. However, GT80 also tried to convert their 
results on torque exerted by the planet to physical (coordinate) 
space, and in doing this they had to resort to certain assumptions. 
In particular, they postulated that the action of each potential 
harmonic, including the deposition of a corresponding torque 
contribution, is localized in the vicinity of a respective 
Lindblad resonance. Under this assumption 
they computed the torque density at each location in physical space 
as the simple product of the spatial density of Lindblad resonances 
$|dk_y/dx|$ and the amplitude of the torque component in Fourier space 
$T_k\equiv\int_{-\infty}^\infty (dT/dx)_k dx$, corresponding to $k_y$
with Lindblad resonance at this particular location (i.e. at 
$k_y=2/(3x)$). Equation 
(\ref{eq:dTdx_GT}) represents the result of such a calculation 
in the limit $x/h\gg 1$. Not surprisingly, one finds 
the sign of $(dT/dx)^{\rm LR}$ to be the same for any $x$ simply 
because the Fourier amplitudes of the torque have the same sign 
independent of $m$ --- there is no disagreement\footnote{Analytical 
calculation of $T_k$ in the limit $k_y h\lesssim 1$ using equation 
(\ref{eq:T_Hk}) and Airy representation (\ref{eq:Imv}), 
(\ref{eq:I}) for $\Im(v)$ results in the expression 
analogous to equation (13) of 
GT80.} on this issue between us and GT80.

What we find in this work is that the Fourier contribution to
the torque density of each azimuthal harmonic of the potential 
$(dT/dx)_k$ is in fact {\it spread out} over a finite portion 
of the disk, rather than localized in a narrow 
region around each Lindblad resonance. In this situation, the 
overlap of many Fourier modes at a given point in space is 
inevitable and the calculation of the full torque density 
in real space represents a rather challenging task, requiring 
direct integration of the equation (\ref{eq:T_Hk}). As a 
result of carrying out this exercise, one finds the 
distribution of the torque density 
different from (\ref{eq:dTdx_GT}) at large values of $x$, 
as clearly demonstrated in Figure \ref{fig:neg_torque}a. 

This reasoning is best illustrated if we temporarily abandon
the shearing sheet approximation and consider the disk-satellite 
interaction in the full cylindrical geometry. Then, as pointed out by
GT80, coupling between the planetary potential and the disk occurs 
at discrete Lindblad resonances radially separated from
the planetary orbit by $x_{L,m}\approx 2r_p/(3m)$, where $m=k_yr_p$
in the azimuthal wavenumber of the potential harmonic and an 
assumption $m\gtrsim 1$ has been made. The distance between the
adjacent $m$-th and $(m+1)$-th resonances is 
$\Delta x_{L,m}=x_{L,m}/m=x_{L,m}(h/r_p)(k_y h)^{-1}$. 

We need to compare this distance to the width 
$\delta x_m$ of the $m$-th resonance. 
According to Airy representation (see \S \ref{subsect:Airy_rep} 
and Appendix \ref{sect:Airy}) the latter is set by the condition 
$t\sim 1$ or $2a^{2/3}(\xi-1)\sim 1$ (valid near the Lindblad 
resonance), see equations (\ref{eq:defs1}) and 
(\ref{eq:near_peak}). One then finds that the resonance width 
$\delta x_m$ is given by $\delta x_m\sim x_{L,m}a^{-2/3}\sim 
x_{L,m}(k_y h)^{2/3}$ for $k_y h\lesssim 1$. Note
that $\delta x_m\ll x_{L,m}$.

A somewhat more illuminating way of deriving $\delta x_m$ is
based on the dispersion relation for the tightly wound density 
waves (Binney \& Tremaine 2008), which is valid for $m\gtrsim 1$
or $k_yh\gtrsim (h/r)$:
\ba
\kappa^2+k_r^2c_s^2=m^2(\Omega-\Omega_p)^2\approx
\frac{4m^2A^2}{r_p^2}x^2,
\label{eq:disp_rel}
\ea
where $k_r$ is the radial wavenumber of the wave. The exact 
resonance location $x_{L,m}=r_pm^{-1}(\kappa/2|A|)$ corresponds 
to the distance where $k_r=0$ and the wave couples best to the 
external potential. As the radial separation from the resonance
increases $k_r$ grows until at $|x-x_{L,m}|\sim\delta x_m$ 
one gets $k_r\delta x_m\sim 1$. Beyond this point the wave 
eigenfunction starts oscillating rapidly, and coupling to 
the external potential becomes less effective. 
Plugging $k_r\sim \delta x_m^{-1}$ and $x=x_{L,m}+\delta x_m$ into 
the dispersion relation (\ref{eq:disp_rel}) and expanding it in
terms of $\delta x_m\ll x_{L,m}$ one finds $\delta x_m\sim 
h(k_yh)^{-1/3}$, the same expression as that as resulting 
from the Airy representation.

Using these estimates of $\Delta x_{L,m}$ and $\delta x_m$ 
one can easily see that $\delta x_m\gtrsim \Delta x_{L,m}$ 
and resonances located at distance $x$ from the perturber 
overlap as long as
\ba
x\lesssim h\left(\frac{r_p}{h}\right)^{3/5}=
r_p\left(\frac{h}{r_p}\right)^{2/5}.
\label{eq:overlap}
\ea
Equivalently, overlap occurs for modes having
\ba
k_yh\gtrsim \left(\frac{h}{r_p}\right)^{3/5} ~~~~\mbox{or}
~~~~m\gtrsim \left(\frac{r_p}{h}\right)^{2/5}. 
\label{eq:overlap_ky}
\ea
Interference of different modes takes place even though the
width of each resonance $\delta x_m$ is small compared to 
the resonance separation $x_{L,m}$ from the perturber's 
orbit.

Thus, whenever $r_p/h\gg 1$, which is always the case for 
protoplanetary and many other kinds of accretion disks, 
resonances overlap in the vicinity of the perturber's 
orbit (for $x$ satisfying equation (\ref{eq:overlap})) and 
the assumption of well-isolated resonances adopted 
by GT80 for their torque density calculation fails.
Given that in our calculation the negative
values of $dT/dx$ appear only for $x>x_-\approx 3.2h$ 
(see \S \ref{subsect:neg_torque}) one needs the
inverse of the disk aspect ratio to be 
$r_p/h\gtrsim (x_-/h)^{5/3}\approx 7$ for the negative 
$dT/dx$ to appear. This requirement is easily fulfilled in
a large variety of astrophysical accretion disks, including 
protoplanetary disks.

In the framework of the shearing sheet approximation, which 
corresponds to the limit $r_p/h\to \infty$, it is not 
surprising, according to equation (\ref{eq:overlap}), that 
resonances overlap for {\it all} $x$ and the isolated resonance 
approximation adopted 
in GT80 never holds. In this limit our asymptotic expression
(\ref{eq:dTdx_full}) works fine for arbitrary $x$.

At the same time, far from the perturber's orbit, at 
separations not obeying the condition (\ref{eq:overlap}) 
resonances become sufficiently spread apart to act in 
isolation. In this case one would expect the GT80's 
expression (\ref{eq:dTdx_GT}) to be valid. Thus, when the full
cylindrical geometry of the disk-planet interaction is 
accounted for $dT/dx$ at fist becomes negative close to 
the planet, for $x>x_-$, but then should change
sign and become positive again (outside of the region defined 
by the inequality (\ref{eq:overlap})) in accordance with the GT80 
calculation. The details of this transition can only be 
elucidated via linear calculation of the disk-planet coupling 
in the cylindrical geometry, which is beyond the scope of 
this work.


\subsection{Comparison with previous work.}
\label{sect:previous}

Quite interestingly, the existence of the negative torque density 
phenomenon could have been inferred from calculations done even 
prior to the work of Dong \etal (2011). In particular, Bate 
\etal (2003) performed global three-dimensional 
simulations of the disk-planet interaction in cylindrical 
geometry. As a part of this calculation they have computed 
torque density produced by planets of different masses. Close 
inspection of their Figure 12 shows that at small masses, 
$M_p\lesssim 0.03$ M$_J$, the torque density changes sign at
a finite separation from the planet. This effect does not disappear
as $M_p$ is reduced strongly suggesting that it is a linear phenomenon.

More recently D'Angelo \& Lubow (2008) have computed radial 
run of $dT/dx$ as one of the by-products of their three-dimensional global 
simulations. In their Figure 7 one
can again see the torque density changing sign for low enough 
$M_p\lesssim 10$ M$_\oplus$, which is in good agreement with the 
torque density behavior that could have been inferred from the 
work of Bate \etal (2003). Similar torque density pattern can 
also be seen in D'Angelo \& Lubow (2010).

Finally, Muto \& Inutsuka (2009) have carried out a local linear 
analysis of interaction between a planet and the surrounding viscous
disk. Even though their results are affected to some extent by
the assumed non-zero value of viscosity (our study assumes viscosity 
to be zero), their torque density still exhibits negative values 
beyond $x\approx 3h$ (see their Figure 5) even at the lowest 
values of viscosity, in agreement with our work.

The most likely reason for why the negative torque density phenomenon 
went unnoticed in previous studies lies in the small amplitude of
this effect. Indeed, the total negative torque contribution obtained by
integrating $dT/dx$ from $x_-\approx 3.2h$ to $\infty$ amounts to
$\lesssim 4\%$ of the total integrated torque 
$T(\infty)=\int_0^\infty(dT/dx)dx$. Thus, it is not surprising that 
this real effect was previously ignored, most likely on the basis of 
it being due to some numerical issues (poor resolution, small 
simulation box size, etc.). 

Nevertheless, the apparent presence of the negative torque phenomenon 
in the aforementioned earlier works provides additional support for 
the robustness of the results presented in this paper. In particular, 
one may wonder whether our conclusions based on a linear calculation 
performed in two-dimensional shearing sheet geometry would hold 
when some of our assumptions --- local approximation, linearity 
of the perturbation, two-dimensional geometry --- are relaxed. In this 
regard we simply point out here that both Bate \etal (2003) and D'Angelo 
\& Lubow (2008) ran three-dimensional, global simulations for 
a variety of planetary masses and one can still discern the 
presence of the negative torque phenomenon in their results. 
This gives us reassurance in the ubiquity of this rather subtle 
but important (see \S \ref{sect:implications}) feature
of the disk-planet interaction even in more sophisticated 
and realistic geometries.


\subsection{Accuracy of the Airy representation.}
\label{sect:acc_Airy}

Another interesting result of our study concerns the accuracy 
with which the analytical Airy approximation described in \S 
\ref{subsect:Airy_rep} and Appendix \ref{sect:Airy} characterizes
the details of the disk-planet interaction. Approximation of
the behavior of the perturbed fluid variables in terms of 
Airy functions has been used by many authors before 
(Artymowicz 1993; Ward 1986) but only 
to represent the behavior in the vicinity of Lindblad 
resonances. As such the Airy representation served only 
as a {\it local} analytical tool for computing the azimuthal 
Fourier components of the density wave properties such as 
the angular momentum flux.

In our study we use Airy representation in a {\it global} sense, to 
describe the spatial variation of the fluid variables for the
values of $x$ not limited to the locations of the Lindblad 
resonances. This procedure works well for describing the 
behavior of the Fourier harmonics of the azimuthal velocity 
perturbation $v$ and of the torque density
$dT/dx$ at {\it any} $x$ as long as the wavenumber of a particular 
mode satisfies the condition $a\gtrsim 1$, which according to
equation (\ref{eq:C2}) is true whenever $k_yh\lesssim 1$ 
or $k_yh\gtrsim 1$. Moreover, for $x\gtrsim h$ this 
approximation is quite reliable even for $k_yh\sim 1$.
As a result the Airy representation successfully reproduces 
the asymptotic behavior of $dT/dx$ and $F_{H,k}$ as a function 
of $x$ in the limit $x/h\gg 1$ as we have shown
in \S \ref{subsect:neg_torque} and \S \ref{sect:flux}. 
And as Figures \ref{fig:dsigma} and \ref{fig:dTdx} demonstrate 
one can also use Airy representation to reproduce with 
decent accuracy the density wave properties in physical space. 

These conclusions are in agreement with 
the results of Heinemann \& Papaloizou (2011) who found that the WKBJ 
approximation valid whenever $a\gtrsim 1$ works very well 
for describing the evolution of the spiral density waves 
excited by the turbulence in the disk.


\section{Astrophysical implications.}
\label{sect:implications}


Understanding the negative torque density phenomenon 
(see \S \ref{subsect:neg_torque}) in uniform disks is the 
major result of our work. It is natural to ask whether this 
effect can somehow change global consequences of the 
disk-satellite interaction in real disks.

It is known that planets, which are not massive enough to 
open a gap around their orbit can migrate radially because 
of the small asymmetry of their gravitational coupling to 
the inner and outer parts of the disk (GT80); this is the 
so-called Type I migration (Ward 1997). In our shearing-sheet 
setup such asymmetry is absent by construction and planets 
do not migrate. But the negative torque density is
still going to be present in real disks (which are close to 
being uniform locally, near the planet) and one may wonder 
whether it may affect the speed of Type I migration. The 
answer is no, because the speed of migration is insensitive 
to the spatial structure of $dT/dx$ and depends only 
on the difference of {\it the full one-sided torques} exerted 
on the inner and outer portions of the disk. As we have shown in 
\S \ref{sect:flux} the one-sided torque is not changed in our 
calculation compared to the GT80 prediction, so the speed of
Type I migration stays the same as long as the disk is roughly 
uniform. 

Deposition of the angular momentum carried by the density waves 
drives evolution of the disk away from its initially uniform 
state. This evolution depends on the spatial pattern of $dT/dr|_d$ 
(see equation (\ref{eq:dTdr_d})) representing the wave angular 
momentum that is being deposited in the disk material. According 
to equation (\ref{eq:dTdr_d}) the latter quantity depends not only 
on $dT/dx$ which describes the {\it driving} of the waves 
but also on the process that leads to their {\it damping} and 
transfer of the wave angular momentum to the disk material
(Takeuchi \etal 1996; Rafikov 2002). However, once the 
latter is understood (i.e. the explicit form of operator 
${\cal L}$ in equation (\ref{eq:dTdr_d}) is determined) the 
disk evolution is going to be fully determined by the dependence 
of $dT/dx$ on $x$, and thus should be directly affected by the 
negative torque density phenomenon. 

The influence of this effect on $dT/dr|_d$ should depend on the 
details of the damping process. Since the negative torque density 
is a rather 
small effect (the full amount of negative torque is just several 
per cent of the full torque, see \S \ref{sect:flux}) it should 
not affect $dT/dr|_d$ if the damping is weak and the characteristic
distance over which wave is dissipated is $l_d\gg h$. What happens
in the case of strong damping, when $l_d$ becomes comparable 
to the distance between individual Lindblad resonances, is not 
so obvious. In particular, it is not clear whether $dT/dr|_d$ should
become negative at $|x|\gtrsim x_-$ as does $dT/dr$, or whether 
$dT/dr|_d$ will tend to converge to the GT80 prediction 
(\ref{eq:dTdx_GT}). This issue can be addressed in the future 
by calculations making explicit assumptions about the damping 
mechanism. And one should keep in mind that strong damping can 
affect not only propagation of the waves but also their 
excitation in a non-trivial way (Muto \& Inutsuka 2009).

It is worth mentioning that, with the rare exceptions of the 
works by Takeuchi \etal (1996) and Rafikov (2002), in the majority 
of studies of the perturber-driven disk evolution the wave 
damping process 
is essentially ignored and it is assumed that the amount 
of angular momentum transferred to the disk material 
is given simply by $dT/dr$ (e.g. Lin \& Papaloizou 1986;
Chang \etal 2010), i.e. the assumption $dT/dr|_d=dT/dr$ 
is usually made. However, the circumstances under which 
such assumption is justified have never been explicitly 
clarified\footnote{It is clear that this assumption is not 
appropriate in the case of weak wave dissipation (Goodman \& 
Rafikov 2001; Rafikov 2002). Whether it may be reasonable in 
the case of strong wave damping remains to be explored.}. 
In general the correct approach to understanding 
the disk evolution due to tidal interaction must incorporate 
{\it both} the correct form of $dT/dr$ (or $dT/dx$) {\it and} 
the description of the wave damping. If the latter is ignored
then according to Goldreich \& Nicholson (1989) the disk 
should not evolve at all. 

Knowledge of the spatial behavior of the torque density $dT/dx$ 
(or $dT/dr$) produced by a perturber is one of the key ingredients  
for understanding the process of gap opening. In particular, 
in steady state the shape of the gap is uniquely determined by 
the balance of the viscous torque density in the disk and the 
radial divergence of the angular momentum flux carried by the 
density waves $dT/dr|_d$.

A proper calculation of the gap shape is important not only 
in itself but also for computing the full torque exerted by
the perturber on the disk. If this torque is non-zero (e.g. 
due to the disk being absent on one side) then the perturber
will migrate (Type II migration in presence of a gap, see Ward (1997)) 
as a result of the angular momentum conservation.
The speed of migration depends on the shape of the gap 
since the surface density profile is one of the ingredients 
determining how much net torque the perturber deposits 
in the disk (Ward 1997).

Calculation of the excitation torque density $dT/dx$ 
in a non-uniform disk must 
be modified compared to that in the constant $\Sigma$ disk.
This is usually accomplished by computing $dT/dx|_{\rm nu}$ in
a non-uniform disk according to the following prescription:
\ba
\frac{dT(x)}{dx}\Bigg|_{\rm nu}=\Sigma(x)
\frac{1}{\Sigma_0}\frac{dT(x)}{dx}\Bigg|_{\rm u},
\label{eq:dTdx_nu}
\ea
where $dT/dx|_{\rm u}$ is the torque density in a uniform 
disk, a quantity which is the subject of the calculations 
done in GT80 and in our work (where we call it just $dT/dx$). 
One normally uses the GT80's
prescription (\ref{eq:dTdx_GT}) for $dT/dx|_{\rm u}$ (Lin \& 
Papaloizou 1986; Trilling \etal 1998; Armitage \& Natarajan 
2002; Armitage \etal 2002; Lodato \etal 2009) or a 
modified version thereof. Given that in this case the 
torque density has a definite sign, the tidal interaction 
between the perturber and the gap edges is repulsive and the
planet drives a {\it positive} angular momentum flux through the 
disk, which is consistent with the gap opening picture.

Our revision of the $dT/dx$ calculation and the discovery 
of the negative torque density phenomenon might be thought 
of as giving rise to an interesting paradox. With our new 
expression (\ref{eq:dTdx_full}) for the torque density one 
may think that equation (\ref{eq:dTdx_nu}) predicts 
$dT/dx|_{\rm nu}<0$ for $x>x_-\approx 3.2 h$. Then the 
application of the prescription (\ref{eq:dTdx_nu}) for 
computing the full torque in a broad gap with width
exceeding $2x_-$ would result in the {\it negative} angular 
momentum flux accumulated by the wave and the {\it attractive} 
interaction between the planet and the fluid at the gap 
edges. Apparently,
under such circumstances the gap would not be able to 
appear in the first place.

The resolution of this apparent paradox lies in the
use of the prescription (\ref{eq:dTdx_nu}) to describe the 
torque density in an inhomogeneous disk. Indeed, the 
calculation of $dT/dx|_{\rm u}$ is based on the solutions of the
perturbed fluid equations (\ref{eq:v})-(\ref{eq:dS})  
derived under the assumption of a {\it uniform} density. 
However, in a non-uniform disk these equations get
modified and new terms caused by the gradients of 
$\Sigma$ appear in them. These terms are especially 
important at the steep edges of gaps or cavities. Their
presence significantly modifies the structure of
the velocity eigenfunctions and makes the prescription
(\ref{eq:dTdx_nu}) irrelevant. In general one should 
not expect equation (\ref{eq:dTdx_nu}) to provide correct 
results even for the excitation torque density $dT/dx$ (not 
mentioning the torque density $dT/dx|_d$ deposited in the 
disk by the density wave).

A fully self-consistent calculation of $dT/dx|_{\rm nu}$ should 
be based on equations explicitly incorporating disk 
non-uniformity and thus has little to do with $dT/dx|_{\rm u}$.
An example of such a self-consistent linear calculation of 
$dT/dx|_{\rm nu}$ is provided in Petrovich \& Rafikov (2012; 
in preparation) where it is shown that a proper account of 
the disk non-uniformity in fluid equations results 
in the positive angular momentum flux produced by the
perturber and the repulsion of fluid away from its
orbit for gaps of arbitrary width.

Knowledge of the behavior of the torque density $dT/dr$ is
also important for determining the surface density 
profile in cavities or gaps formed in accretion disks by 
embedded binary SMBHs. In the case of extreme mass 
ratio inspirals the gap formed by the lower mass BH
can be quite narrow (Chang \etal 2010) so that the results 
obtained in the local shearing sheet approximation 
remain valid. In particular, the negative torque density 
phenomenon may be relevant for this type of objects 
if the mass of the secondary is insufficient to 
significantly perturb the disk surface density and 
open a gap. 

However, in the case of SMBHs of comparable mass the binary 
usually clears out a large cavity with inner radius comparable 
to the binary separation in the disk around itself 
(MacFadyen \& Milosavljevi\'c 2008; Shi \etal 2011). In that case 
understanding the torque density distribution requires one 
to carry out a global calculation of the density wave 
excitation by the binary
in a highly non-uniform disk in full cylindrical geometry.  
But even in this much more complicated setting one may 
still use the intuition 
gained in our present study. In particular, equation 
(\ref{eq:overlap}) suggests that the negative torque 
phenomenon is not going to be important since the 
range of $x$ where the resonances overlap falls inside
the cavity. At the inner edge of the cavity, at separations 
of order the binary semi-major axis the resonances are well 
isolated and procedure adopted in GT80 to compute $dT/dr$ may
actually work. 

On the other hand, if the cavity is very 
wide\footnote{For equal mass ratio
SMBH binary MacFadyen \& Milosavljevi\'c (2008) found the cavity
size to be about twice the binary semi-major axis.} then the 
torque is generated by only a handful of individual 
modes since only a small number of low-$m$ Lindblad resonances 
are going to be located in the region with appreciable 
gas surface density. As a result, the pattern of $dT/dx$ may be more 
reminiscent of the $(dT/dx)_k$ corresponding to a single 
azimuthal Fourier harmonic of the potential, which usually 
exhibits complicated oscillatory pattern, see upper right panel 
in Figure \ref{fig:dTdx_k} and KP93. This expectation agrees well with 
the results of numerical simulations of MacFadyen \& Milosavljevi\'c 
(2008), Cuadra \etal (2009), Farris \etal (2011), 
and Shi \etal (2011) who have found oscillating $dT/dx$ 
in disks with large cavities cleared by the SMBH binaries.


\section{Summary.}
\label{sect:summary}


We have carried out linear calculation of the gravitational 
coupling between the uniform gaseous disk and an embedded massive 
perturber. Our work goes beyond similar earlier studies by
putting emphasis on understanding the behavior of the 
perturbed fluid variables in real space as opposed to Fourier
space. This allows us to address the nature of the recently 
discovered negative torque phenomenon and explain its origin 
fully in the framework of linear theory. 

We come up with the global analytical representation of the 
spatial variation of the perturbed fluid variables in terms 
of the Airy functions, which is shown to work quite well for
describing the properties of the planet-generated density waves 
both in Fourier and in real space. Using this approximation
we derive the asymptotic behavior of the torque density far
from the perturber and show that it matches the results of 
numerical calculations. Our linear calculations 
confirm that the torque density changes sign at a finite
separation from the perturber in full agreement with the 
results of the direct hydrodynamical simulations. This 
provides an important correction to the calculations of GT80.

These results have broad implications for understanding 
tidal interaction of planets with circumstellar disks and 
of binary SMBHs with the circumbinary accretion
disks. In particular, our calculations provide a pathway 
for deriving the spatial torque distributions in such disks, 
which is important for understanding the issues of gap 
opening by planets and gas clearing by binary SMBHs.


\acknowledgements 

We are grateful to Jeremy Goodman for useful discussions, the 
referee Peter Goldreich for constructive criticisms,  
and Ruobing Dong for sharing his simulation data with us. 
RRR thanks the Lebedev Physical Institute 
and the Institute of Space Research (Moscow) for hospitality 
during the time when part of this work was performed.
The financial support for this work is provided by 
the Sloan Foundation, NASA grant NNX08AH87G, NSF grant 
AST-0908269, and CONICYT Bicentennial Becas-Chile 
fellowship awarded to CP.

\appendix


\section{Airy representation.}  
\label{sect:Airy}


The following results extensively use the relations presented 
in Chapter 19 of Abramowitz \& Stegun (1972; hereafter AS). 
Homogeneous parabolic cylinder equation
\ba
\frac{\partial^2 v}{\partial X^2}+v\left(\frac{X^2}{4}-a\right)=0,
\label{eq:homog}
\ea
i.e. the equation (\ref{eq:parab}) with $R$ set to zero has 
two independent homogeneous solutions, which behave 
as purely outgoing waves at infinity (Narayan \etal 1987):   
\ba
&& v_+=E(a,X)=\rmk^{-1/2}W(a,X)+i\rmk^{1/2}W(a,-X),
\nonumber\\
&& v_-=E^\star(a,-X)=\rmk^{-1/2}W(a,-X)-i\rmk^{1/2}W(a,X),
\label{eq:Esols}\\
&& \rmk\equiv \sqrt{1+e^{2\pi a}}-e^{\pi a},
\label{eq:k}
\ea
where $W(a,X)$ is the standard solution of equation (\ref{eq:homog}),
see chapter 19.17 of AS. One can easily show that the Wronskian
$\Delta=W\left\{E(a,X),E^\star(a,-X)\right\}=2e^{\pi a}$ is 
independent of $X$.

In chapter 19.20 of AS it is shown that when $a$ is large and positive
($a\gg 1$) the parabolic cylinder function $W(a,X)$ can be expressed
via Airy functions for any $X>0$ in the following way:
\ba
W(a,X)\sim\sqrt{\pi}\frac{e^{-\pi a/2}}{(4a)^{1/4}}
g(X)\Bi(-t),~~~~~~
W(a,-X)\sim2\sqrt{\pi}\frac{e^{\pi a/2}}{(4a)^{1/4}}
g(X)\Ai(-t),
\label{eq:WviaAiry}
\ea
where 
\ba
\xi(X)& \equiv &\frac{X}{2\sqrt{a}},~~~~~t(\xi)\equiv (4a)^{2/3}\tau(\xi),
~~~~~g(X)\equiv\left[\frac{t(\xi)}{\xi^2-1}\right]^{1/4}
\label{eq:defs1}\\
\tau(\xi) &=& -\left(\frac{3}{8}\right)^{2/3}\left(\mbox{arccos}~\xi-
\xi\sqrt{1-\xi^2}\right)^{2/3},~~~~~~\xi\le 1,\nonumber\\
&=& \left(\frac{3}{8}\right)^{2/3}\left(\xi\sqrt{\xi^2-1}-
\mbox{arccosh}~\xi\right)^{2/3},~~~~~~\xi\ge 1.
\label{eq:tau}
\ea

We will also need to know the asymptotic behavior of Airy functions
as $X\to\infty$ (here $\zeta\equiv (2/3)X^{3/2}$):
\ba
&&\Ai(X)\sim\frac{e^{-\zeta}}{2\sqrt{\pi}X^{1/4}},~~~~~
\Bi(X)\sim \frac{e^\zeta}{\sqrt{\pi}X^{1/4}},
\nonumber\\
&&\Ai(-X)\sim\frac{\cos(\zeta-\pi/4)}{\sqrt{\pi}X^{1/4}},~~~~~
\Bi(-X)\sim -\frac{\sin(\zeta-\pi/4)}{\sqrt{\pi}X^{1/4}}.
\label{eq:asymp}
%
\ea 

Full solution of inhomogeneous equation (\ref{eq:parab}) 
behaving as purely outgoing wave for $x\to \pm\infty$ is
expressed via the homogeneous solutions $v_\pm$ as
\ba
v(k_y,x)=-\frac{v_+}{\Delta}\int\limits_{-\infty}^z
v_-(s)R(s)ds-\frac{v_-}{\Delta}\int\limits^{\infty}_z
v_+(s)R(s)ds.
\label{eq:inhom_sol}
\ea
Here it is understood that $z=z(x)$ and $R(s)=R(x(s))$, where 
the corresponding dependencies are given by equations (\ref{eq:var_trans})
and (\ref{eq:R}).

Now we consider the behavior of the solutions for positive $x$ (and $z$).
For $a\gtrsim 1$ and $z>0$ one can write using equations 
(\ref{eq:Esols})-(\ref{eq:WviaAiry})
\ba
v_+  &\sim & \sqrt{2}e^{\pi a/2}W(a,z)+i
\frac{e^{-\pi a/2}}{\sqrt{2}}W(a,-z)\sim
\frac{\sqrt{\pi}}{a^{1/4}}g(z)\left[\Bi(-t(z))
+i\Ai(-t(z))\right],
\label{eq:v+pos}\\
v_- &\sim & \sqrt{2}e^{\pi a/2}W(a,-z)-i
\frac{e^{-\pi a/2}}{\sqrt{2}}W(a,z)
\nonumber\\
&\sim &
\frac{\sqrt{\pi}}{a^{1/4}}g(z)
\left[2e^{\pi a}\Ai(-t(z))
-i\frac{e^{-\pi a}}{2}\Bi(-t(z))\right],
\label{eq:v-pos}
\ea
where $\xi=z/(2\sqrt{a})$ and $t(z)$ implies $t$ calculated 
with this value of $\xi$.

Integrals in equation (\ref{eq:inhom_sol}) involve integration 
over both positive and negative $z$ (or $x$), so we need 
the Airy function approximation for $v_\pm$ for 
negative arguments as well:
\ba
v_+ &\sim & \sqrt{2}e^{\pi a/2}W(a,-|z|)+i
\frac{e^{-\pi a/2}}{\sqrt{2}}W(a,|z|)
\nonumber\\
&\sim &
\frac{\sqrt{\pi}}{a^{1/4}}g(-z)
\left[2e^{\pi a}\Ai(-t(-z))
+i\frac{e^{-\pi a}}{2}\Bi(-t(-z))\right],
\label{eq:v+}\\
 v_- &\sim & \sqrt{2}e^{\pi a/2}W(a,|z|)-i
\frac{e^{-\pi a/2}}{\sqrt{2}}W(a,-|z|)
\nonumber\\
&\sim &
\frac{\sqrt{\pi}}{a^{1/4}}g(-z)\left[\Bi(-t(-z))
-i\Ai(-t(-z))\right],
\label{eq:v-}
\ea
where now $z<0$, $\xi=-z/(2\sqrt{a})$, and $t(-z)$ means
$t$ computed with this value of $\xi$. 

Plugging the expressions (\ref{eq:v+})-(\ref{eq:v-pos}) 
into equation (\ref{eq:inhom_sol}) we find that for $a\gg 1$ 
and $z>0$ 
\ba
\Re(v) &\sim & -\frac{\pi g(z)}{a^{1/2}}\left\{
\Ai(-t(z))\left[\int\limits^{\infty}_z g(s)\Bi(-t(s))R(s)ds
+\frac{e^{-2\pi a}}{4}\int\limits^z_0 g(s)\Bi(-t(s))R(s)ds
\right.\right.
\nonumber\\
&-& \left.\left.
\frac{e^{-\pi a}}{2}\int\limits^{\infty}_0 g(s)\Ai(-t(s))R(s)ds
\right]+
\Bi(-t(z))\left[\int\limits_0^z g(s)\Ai(-t(s))R(s)ds
\right.\right.
\nonumber\\
&+& \left.\left.
\frac{e^{-2\pi a}}{4}\int\limits_z^\infty g(s)\Ai(-t(s))R(s)ds
-\frac{e^{-\pi a}}{2}\int\limits^{\infty}_0 g(s)\Bi(-t(s))R(s)ds
\right]\right\},
\label{eq:Rev_full}\\
\Im(v) &\sim & -\frac{\pi g(z)}{a^{1/2}}
\left[\Ai(-t(z))+\frac{e^{-\pi a}}{2}\Bi(-t(z))\right]
\nonumber\\
&\times &
\int\limits_0^{\infty} g(s)\left[\Ai(-t(s))-\frac{e^{-\pi a}}{2}
\Bi(-t(s))\right]R(s)ds.
\label{eq:Imv_full}
\ea
In deriving these expressions we have used the fact that 
\ba
\int_0^\infty g(s)\left\{\Ai,\Bi\right\}(-t(s))R(s)ds=
-\int^0_{-\infty} g(-s)\left\{\Ai,\Bi\right\}(-t(-s))R(s)ds
\nonumber
\ea
because 
$R(-z)=-R(z)$.

These formulae contain a number of terms which are exponentially
small in the limit $a\gg 1$.
Indeed, using the relations (\ref{eq:asymp}) and definitions 
(\ref{eq:defs1})-(\ref{eq:tau}) one can easily show that even 
for $z=\xi=0$ when $\zeta=\pi a/2$ is largest and 
the term $\propto \Bi$ is 
exponentially large (and the term $\propto \Ai$ is 
exponentially small), terms like $e^{-\pi a}\Bi$ are still 
subdominant as $O\left(e^{-\pi a/2}\right)$. All other terms 
containing exponential factors are even more subdominant.
Thus, we can safely neglect such terms for $a\gg 1$ and $z>0$ 
to get an approximate form (\ref{eq:Rev})-(\ref{eq:I-def}) 
of the solution for $v$.

We now evaluate the integral $I_+$ appearing in equation
(\ref{eq:Rev}) for a Keplerian disk 
in the limit $k_yh\ll 1$ when $a\approx (3k_yh)^{-1}\gg 1$, with 
$R(s)=R\left(h(3k_yh)^{-1/2}s\right)$. 
The second term in this integral is exponentially small when 
$a\gg 1$ and can be dropped for simplicity. 
The asymptotic behavior (\ref{eq:asymp}) 
of Ai suggests that the rest of the integral is dominated by $s$ such
that $t(s)\sim 1$, since for other $s$ the integrand is either
exponentially small or rapidly oscillates. From the definition
(\ref{eq:defs1}) of $t$ one can see that this requires 
$\tau\ll 1$, resulting in integral being dominated by 
the small vicinity ($|\xi-1|\ll 1$) of 
$s=2\sqrt{a}\approx 2(3k_yh)^{-1/2}\gtrsim 1$. 
One can easily show that in this region 
\ba
t\approx 2a^{2/3}(\xi-1)\left(1+\frac{\xi-1}{10}\right),~~~~~
g\approx a^{1/6}\left(1-\frac{\xi-1}{10}\right). 
\label{eq:near_peak}
\ea
Using the fact that $\int_{-\infty}^\infty
Ai(s)ds=1$ and that the driving term $R(s)$ varies only
weakly in the region where $|\xi-1|\ll 1$ we find 
for $x<0$ (retaining in (\ref{eq:near_peak}) only the terms 
of the lowest order in $\xi-1$) that $I_+$ is given by 
equation (\ref{eq:I}).


\section{Torque density.}
\label{sect:T_H}

We now proceed to calculate the torque density $dT/dx$. Using 
equations (\ref{eq:Imv}), (\ref{eq:I-def}), \& (\ref{eq:T_Hk}) 
we write
\ba
\left(\frac{dT}{dx}\right)_k=\frac{8\pi\Sigma_0k_y}{k_y^2c^2+4B^2}
\phi\frac{\pi I_+}{a^{1/2}}\left\{B\frac{\partial}{\partial x}
\left[g(z)\Ai(-t(z))\right]-Ak_y^2x g(z)\Ai(-t(z)) \right\}.
\label{eq:T_Hk_withAiry}
\ea
This expression is accurate up to the exponentially small terms.

To perform differentiation w.r.t. $x$ in this equation we
use the following relations, which can be easily derived 
from equations (\ref{eq:defs1})-(\ref{eq:tau}):
\ba
&&\frac{\partial\tau}{\partial\xi}=\frac{1}{2}\sqrt{\frac{\xi^2-1}{\tau}},
~~~~\frac{\partial t(z)}{\partial x}=\sqrt{a}\frac{(3k_yh)^{1/2}}{h}
(g(z))^{-2},
\label{eq:rels1}\\
&&\frac{\partial g(z)}{\partial x}=-
\sqrt{a}\frac{(3k_yh)^{1/2}}{h}\frac{g(z)}{4(\xi^2-1)}
\left[\frac{\xi}{a}-(g(z))^{-6}\right].
\label{eq:rels2}
\ea
Then we find for a Keplerian disk
\ba
\left(\frac{dT}{dx}\right)_k &=& 
\frac{2\pi^2\Sigma_0\Omega^2 k_y}{k_y^2c^2+\Omega^2/4}
\frac{I_+\phi}{a^{1/2}c}\Bigg\{3(k_yh)(k_yx) g(z)\Ai(-t(z))\nonumber\\
&-&\frac{\sqrt{1+(k_yh)^2}}{g(z)}
\left[\frac{(g(z))^2}{4a}\frac{\xi-a(g(z))^{-6}}{\xi^2-1}\Ai(-t(z))
+\Ai^\prime(-t(z))\right]\Bigg\},
\label{eq:T_Hk_withAiry1}
\ea
where $\Ai^\prime(X)\equiv\partial \Ai(X)/\partial X$. 
This equation is valid for any $x$ (or $z$) as long as $a\gtrsim 1$.

To get the full torque density one needs to substitute 
$(dT/dx)_k$ into equation (\ref{eq:T_H}) and perform integration 
over $k_y$. In general one cannot expect the expression
(\ref{eq:T_Hk_withAiry1}) to represent $(dT/dx)_k$ over the full
range of $k_y$ since it is valid only for $a\gtrsim 1$ and
formally fails for $k_yh\sim 1$. 

However, this is not a problem for obtaining the asymptotic 
expression for the torque density $dT/dx$ far from the planet, 
at $x/h\gg 1$, which we do next. Indeed, $(dT/dx)_k$ is
always proportional to the planetary potential $\phi$, which 
according to equation (\ref{eq:pot}) behaves as $\phi\propto \exp
\left[-(k_yh)(x/h)\right]\ll 1$ for $x/h\gg 1$ and $k_yh\sim 1$  
(this follows from the asymptotic behavior of the modified Bessel 
functions at large values of the argument). As a result, the 
contribution of harmonics with $k_yh\sim 1$, for which $a\sim 1$ 
and analytical expression (\ref{eq:T_Hk_withAiry1}) fails, is 
exponentially suppressed and we need not worry about them. 
For the same reason one can safely neglect the contribution 
to $dT/dx$ produced by modes with $k_yh\gtrsim 1$ even though 
equation (\ref{eq:T_Hk_withAiry1}) works well for them.

We thus expect that for $x/h\gtrsim 1$ torque density $dT/dx$ is 
contributed mainly by modes with $k_yh\lesssim 1$ for which
$\phi$ is not exponentially suppressed, 
$a\approx (3k_yh)^{-1}\gtrsim 1$ and the expression 
(\ref{eq:T_Hk_withAiry1}) can be used (this also allows us to 
neglect a number of terms $O(k_y^2h^2)$ in (\ref{eq:T_Hk_withAiry1})). 
This expression contains terms 
proportional to either $\Ai$ or $\Ai^\prime$ and we look 
at their corresponding integrals $(dT/dx)_1^{as}$ and $(dT/dx)_2^{as}$ 
separately, starting with the former.

Airy function $\Ai(X)$ rapidly oscillates for $X\to -\infty$
and decays exponentially for $X\to \infty$, see equation 
(\ref{eq:asymp}). In both limits 
the corresponding contributions to the integral are small and the 
main contribution is provided by the interval of $k_y$ in which 
$X=-t(z)\sim 1$. From equation (\ref{eq:defs1})
and the fact that $a\gtrsim 1$ it follows that $t\sim 1$ requires 
$\tau(\xi)\ll 1$, and $|\xi-1|\ll 1$. Then according to equation 
(\ref{eq:near_peak}) $-t(z)\to 2a^{2/3}(1-\xi)=
a^{2/3}(2-3k_yx)$ and it is clear that $(dT/dx)_1^{as}$ is dominated
by a narrow vicinity of $k_y=(2/3)x^{-1}$ (with absolute width of 
$\delta k_y\sim |x|^{-1}a^{-2/3}\sim |x|^{-1}(h/|x|)^{2/3}$ and relative 
width $\delta k_y/k_y\sim (h/|x|)^{2/3}\ll 1$). Inside this narrow 
range of $k_y$ all other factors in the integrand function  
vary only weakly and can be evaluated for 
$k_y\approx (2/3)x^{-1}$ and $g$ given by (\ref{eq:near_peak}), and 
taken out of the integral. As a result one finds that
\ba
\left(\frac{dT}{dx}\right)_1^{as}\to \frac{296}{405}
K_0\left(\frac{2}{3}\right)
\left[2K_0\left(\frac{2}{3}\right)+K_1\left(\frac{2}{3}\right)\right]
\frac{(GM_p)^2\Sigma_0}{\Omega^2}\frac{1}{x^4},
\label{eq:dTdx_1}
\ea
where equations (\ref{eq:pot}) and (\ref{eq:I}) were used to 
evaluate $\phi$ and $I_+$.

Evaluation of the torque contribution $(dT/dx)_2^{as}$ proportional to
$\Ai^\prime$ is a more intricate exercise. It is accomplished
via integration by parts, which again results in the integral
of a rather complicated function multiplied by $\Ai(-t(z))$. 
After laborious but straightforward calculation this function 
can be evaluated at $k_y\approx (2/3)x^{-1}$ and the 
whole integral becomes
\ba
\left(\frac{dT}{dx}\right)_2^{as}\to -\frac{16}{27}
\left[\frac{29}{10}K_0\left(\frac{2}{3}\right)-\frac{2}{3}
K_1\left(\frac{2}{3}\right)\right]
\left[2K_0\left(\frac{2}{3}\right)+K_1\left(\frac{2}{3}\right)\right]
\frac{(GM_p)^2\Sigma_0}{\Omega^2}\frac{1}{x^4}.
\label{eq:dTdx_2}
\ea
Summing up (\ref{eq:dTdx_1}) and (\ref{eq:dTdx_2}) we arrive 
at the final asymptotic expression (\ref{eq:dTdx_full}) for 
the torque density valid in the limit $x/h\gg 1$.


\section{Angular momentum flux.}
\label{sect:F_H}

Plugging in (\ref{eq:Imv}), (\ref{eq:I-def}) and 
(\ref{eq:Rev_far}) into (\ref{eq:F_Hk_far}) we obtain
\ba
F_{H,k}&=&\frac{4\pi\Sigma_0k_yc^2}{k_y^2c^2+4B^2}
\nonumber\\
&\times &
\frac{\pi^2I_+^2g(-z)}{a}\left\{
\Ai(-t(z))\frac{\partial}{\partial x}\left[g(z)\Bi(-t(z))\right]-
\Bi(-t(z))\frac{\partial}{\partial x}\left[g(z)\Ai(-t(z))\right]
\right\}.
\label{eq:F_Hk_far1}
\ea

As $x\to \infty$ so does $z\to \infty$ for any $k_y$. 
Then, according to equations
(\ref{eq:defs1})-(\ref{eq:tau}) one has $\xi\to \infty$, 
$\tau(\xi)\to (3\xi^2/8)^{2/3}$, $t(z)\to (3a\xi^2/2)^{2/3}$,
and $g\to (3a/2\xi)^{1/6}$.
Also, in equation (\ref{eq:F_Hk_far1}) we can replace Airy 
functions with their asymptotic behavior (\ref{eq:asymp}) 
in terms of trigonometric functions. Plugging all that into the
equation (\ref{eq:F_Hk_far1}), taking the limit $k_y h\lesssim 1$, 
and using the expression (\ref{eq:I}) 
for $I_+$ one finds for a Keplerian disk 
\ba
F_{H,k}(x\to \infty)=12k_y^2\frac{(GM_p)^2\Sigma_0}{\Omega^2}
\left[wK_0\left(|w|\right)+\frac{1}{3}
K_1\left(|w|\right)\right]^2,
\label{eq:al_fin}
\ea
with $w$ defined in equation (\ref{eq:I}). In the long wavelength 
limit $k_yh\lesssim 1$ one has $w\to 2/3$ and equation (\ref{eq:al_fin})
reduces to equation (\ref{eq:fin}).



 

\end{document}